\documentclass[
reprint,
superscriptaddress,
amsmath,
amssymb,
aps,
prl,
]{revtex4-2}

\usepackage{graphicx}
\usepackage{dcolumn}
\usepackage{bm}
\usepackage{kotex}
\usepackage{verbatim}

\begin{document}
\preprint{APS/123-QED}

\title{Order--Disorder Tricriticality in \(\mathrm{A}_n \mathrm{B}_n\) Star Polymer Melts}

\author{Minhoon Kim}
\affiliation{Department of Physics, Ulsan National Institute of Science and Technology (UNIST), Ulsan 44919, Republic of Korea}
\author{Wonjun Kang}
\affiliation{Department of Physics, Ulsan National Institute of Science and Technology (UNIST), Ulsan 44919, Republic of Korea}
\author{Daeseong Yong}
\affiliation{Center for AI and Natural Sciences, Korea Institute for Advanced Study, Seoul 02455, Republic of Korea}
\author{Junhan Cho}
\affiliation{Department of Polymer Science and Engineering, Dankook University, 152 Jukjeon-ro, Yongin, Gyeonggi-do 16890, Republic of Korea}
\author{Jaeup U. Kim}
\email{jukim@unist.ac.kr}
\affiliation{Department of Physics, Ulsan National Institute of Science and Technology (UNIST), Ulsan 44919, Republic of Korea}

\date{\today}

\begin{abstract}
Tricriticality usually requires tuning an additional thermodynamic parameter. Here we show that, in symmetric $\mathrm{A}_n\mathrm{B}_n$ star-polymer melts, the arm number $n$ itself plays this role and drives the order--disorder transition (ODT) from second order to first order. By developing a sixth-order free-energy expansion within the random phase approximation and comparing it with self-consistent field theory (SCFT) calculations, we analytically identify a tricritical arm number, $n_{\mathrm{tc}}\approx 5.4475$. For $n<n_{\mathrm{tc}}$, the lamellar ordering transition remains continuous and occurs at the spinodal point, $(\chi N)_{\mathrm{s}}\approx 10.495$. For $n>n_{\mathrm{tc}}$, the transition becomes first order, and $(\chi N)_{\mathrm{ODT}}$ shifts below $(\chi N)_{\mathrm{s}}$ with a quadratic dependence near the tricritical point. SCFT calculations confirm the predicted transition character and phase-boundary shift. The origin of this behavior is traced to inter-arm correlations generated by the common junction. We further show that the noninteger tricritical arm number can be effectively realized in binary mixtures of star polymers. This provides a rare analytically tractable example of architecture-induced tricriticality in a microphase-separating polymer system.
\end{abstract}

\maketitle 

{\it Introduction}--- Tricritical behavior occurs at the point where a line of continuous second-order transitions terminates and a first-order transition begins.
It has been extensively studied across various physical systems, including $^{3}\mathrm{He}$-$^{4}\mathrm{He}$ mixtures \cite{Griffiths1970, Kincaid1975}, superconductors \cite{Halperin1974, Mo2002}, liquid crystals \cite{Keyes1973, Alben1973}, and polymer systems \cite{Broseta1990,Holyst1992,Cho2021,Willis2024, Xie2025}.

At the mean-field level, this rich tricritical behavior can be understood within the canonical Landau free-energy framework \cite{Landau1937}:
\begin{equation}
 L(\psi)=a\psi^2 + b\psi^4+c\psi^6.
\end{equation}
Here, $\psi$ is the order parameter, and $c>0$ ensures global thermodynamic stability. The nature of the transition is governed primarily by the signs of the quadratic and quartic coefficients. For $b>0$, the transition occurs at $a=0$ and is continuous, i.e., second order. When the quartic coefficient becomes negative, the transition instead becomes first order. The tricritical point marks the boundary between these two regimes and is defined by $a=b=0$.

Block copolymers (BCPs) \cite{Bates1990} exhibit rich phase behavior arising from the microphase separation of chemically distinct blocks. 
In the long-chain limit, their behavior is commonly described by the Gaussian chain model \cite{Matsen2002}, which provides a minimal coarse-grained framework for such phenomena. Within this framework, the phase behavior of AB BCPs is governed primarily by the composition fraction $f$ and the effective segregation strength $\chi N$, where $N$ is the degree of polymerization and $\chi$ is the Flory–Huggins interaction parameter between A and B segments.

A variety of analytical and numerical approaches have been developed to describe the phase behavior of BCPs at the mean-field level.
On the analytical side, the order–disorder transition (ODT) is naturally treated within a weak-segregation framework because composition modulations remain small near the transition. In this regime, the random phase approximation (RPA) provides a natural basis for a Landau-type expansion of the free energy. Within this framework, Leibler \cite{Leibler1980} developed a pioneering analytical theory for AB diblock copolymers by carrying the expansion to fourth order. In particular, he showed that the symmetric diblock copolymer ($f=1/2$) undergoes a second-order ODT at $(\chi N)_\text{ODT} \approx 10.495$.

Since Leibler's seminal work, second-order RPA analyses~\cite{delacruz1986, Benoit1988} and fourth-order RPA-based Landau expansions~\cite{Mayes1989,Dobrynin1991,Dobrynin1993,Morozov2001} have been widely used to study complex polymer architectures. In particular, Olvera de la Cruz and Sanchez~\cite{delacruz1986} used a second-order RPA calculation for symmetric $\mathrm{A}_n\mathrm{B}_n$ star polymers and obtained an arm-number-independent quadratic instability at $(\chi N)_s\approx 10.495$. Within a second-order treatment, this instability is identified with the ODT. Dobrynin and Erukhimovich~\cite{Dobrynin1991, Dobrynin1993} later developed a computational scheme to evaluate the free energy of $\mathrm{A}_n\mathrm{B}_n$ star polymers through fourth order. They found that, near $f=1/2$, the quartic coefficient becomes negative for $n>5$, indicating that the spinodal need not coincide with the ODT and that a fourth-order truncation is insufficient to determine the transition. As they noted, a complete analysis requires carrying the expansion through sixth order.

The rapidly increasing complexity of higher-order calculations makes sixth-order RPA-based Landau expansions highly challenging. Although several studies addressed aspects of the problem, including diagrammatic schemes for generating higher-order vertex functions~\cite{Aliev2009} and an analysis illustrating the substantial effect of sextic contributions on phase diagrams~\cite{Kuchanov2006}, explicit applications of sixth-order RPA-based free-energy expansions to identify phenomena such as tricritical behavior have remained limited.

Complementary to these analytical developments, self-consistent field theory (SCFT) provides quantitative predictions for BCP phase behavior beyond the weak-segregation regime \cite{Matsen1994, Drolet1999}. Matsen and Gardiner \cite{Matsen2000} first applied SCFT to $\mathrm{A}_n \mathrm{B}_n$ star polymers. More recent studies examined domain spacing for $n\le 5$~\cite{Ji2020} and constructed a partial phase diagram of $\mathrm{A}_{5}\mathrm{B}_{5}$ in the spherical-phase regime~\cite{Li2021}.
However, these studies have been limited to relatively small arm numbers, leaving the large $n$ regime unexplored.

In this work, we use an analytical sixth-order RPA-based theory and SCFT calculations to demonstrate that $\mathrm{A}_n\mathrm{B}_n$ star polymers exhibit rich phase-transition behavior, with the arm number $n$ serving as a tuning parameter. SCFT calculations reveal that the ODT becomes first order for integer arm numbers $n\ge 6$, with $(\chi N)_{\mathrm{ODT}}$ decreasing as $n$ increases. We then construct a systematic sixth-order RPA-based free-energy expansion that analytically identifies the tricritical arm number, yielding $n_{\mathrm{tc}}\approx 5.4475$, and quantitatively describes the transition behavior near this point, in excellent agreement with SCFT. This tricriticality is distinct from the more familiar polymer tricriticality associated with multicritical phase diagrams typically involving a zero-wavenumber macrophase-separation mode~\cite{Broseta1990,Holyst1992,Cho2021,Willis2024,Xie2025}. In this work, by contrast, the order of the ODT changes within the same finite-wavenumber ordering channel, at a fixed \(k^*>0\).

{\it Theory and Results} --- We consider an incompressible melt of $n_p$ $\mathrm{A}_n\mathrm{B}_n$ star polymers, each consisting of $n$ A arms with $fN$ segments and $n$ B arms with $(1-f)N$ segments. Each segment occupies a volume $1/\rho_0$, so the total system volume is $V = N n n_p/\rho_0$. For the field-theoretic description of this system, let $q(\mathbf{r},s)$ and $p(\mathbf{r},s)$ denote the partial partition functions of an A arm and a B arm, respectively, measured from their free ends. They satisfy the following modified diffusion equations (MDEs):
\begin{equation}
\begin{aligned}
\frac{\partial q(\mathbf{r},s)}{\partial s} &= \frac{a^2 N}{6} \nabla^2 q(\mathbf{r},s) - W_\mathrm{A}(\mathbf{r}) q(\mathbf{r},s), \\
\frac{\partial p(\mathbf{r},s)}{\partial s} &= \frac{a^2 N}{6} \nabla^2 p(\mathbf{r},s) - W_\mathrm{B}(\mathbf{r}) p(\mathbf{r},s),
\end{aligned}
\end{equation}
with initial conditions $q(\mathbf{r},0) = p(\mathbf{r}, 0) = 1$. Here, $s$ denotes the position along the contour of the chain, spanning $0 \le s \le f$ for $q(\mathbf{r}, s)$ and $0 \le s \le 1-f$ for $p(\mathbf{r}, s)$. Also, $a$ is the statistical segment length, and $W_\mathrm{A}(\mathbf{r})$ and $W_\mathrm{B}(\mathbf{r})$ are the potential fields acting on the A and B segments, respectively. The total partition function is conveniently evaluated at the junction point as
\begin{equation}
\begin{aligned}
Q = \frac{1}{V} \int_V q^n(\mathbf{r},f)\,p^n(\mathbf{r},1-f)\,d\boldsymbol{\mathrm{r}}.
\end{aligned}
\end{equation}

Let $F$ denote the total free energy of the $\mathrm{A}_n\mathrm{B}_n$ star-polymer system, and define the dimensionless free energy as $\mathcal{F} \equiv F/(n n_p k_{\rm B} T)$. This normalization corresponds to the free energy per AB arm pair, allowing direct comparison across different values of $n$. The dimensionless free energy relative to the disordered phase is written as \cite{Matsen2021} 
\begin{equation}
\begin{split}
{\mathcal F} = -\frac{1}{n}\ln Q + \frac{1}{V} \int_V \left(\frac{W_-^2(\boldsymbol{\mathrm{r}})}
{\chi N}-W_+(\boldsymbol{\mathrm{r}}) \right)d \boldsymbol{\mathrm{r}}  \\
+ \chi N\left(f-\frac12\right)^2 .
\end{split}
\end{equation}
Here, $W_+(\mathbf{r}) =\left( W_\mathrm{A}(\mathbf{r}) + W_\mathrm{B}(\mathbf{r}) \right)/2$ acts as a pressure field that enforces the incompressibility constraint, and $W_-(\mathbf{r}) = \left( W_\mathrm{A}(\mathbf{r}) - W_\mathrm{B}(\mathbf{r}) \right)/2$ is an exchange field that drives A/B segregation. With this definition, the free energy of the disordered phase always vanishes.

\begin{figure} 
\centering
\includegraphics[width=\linewidth]{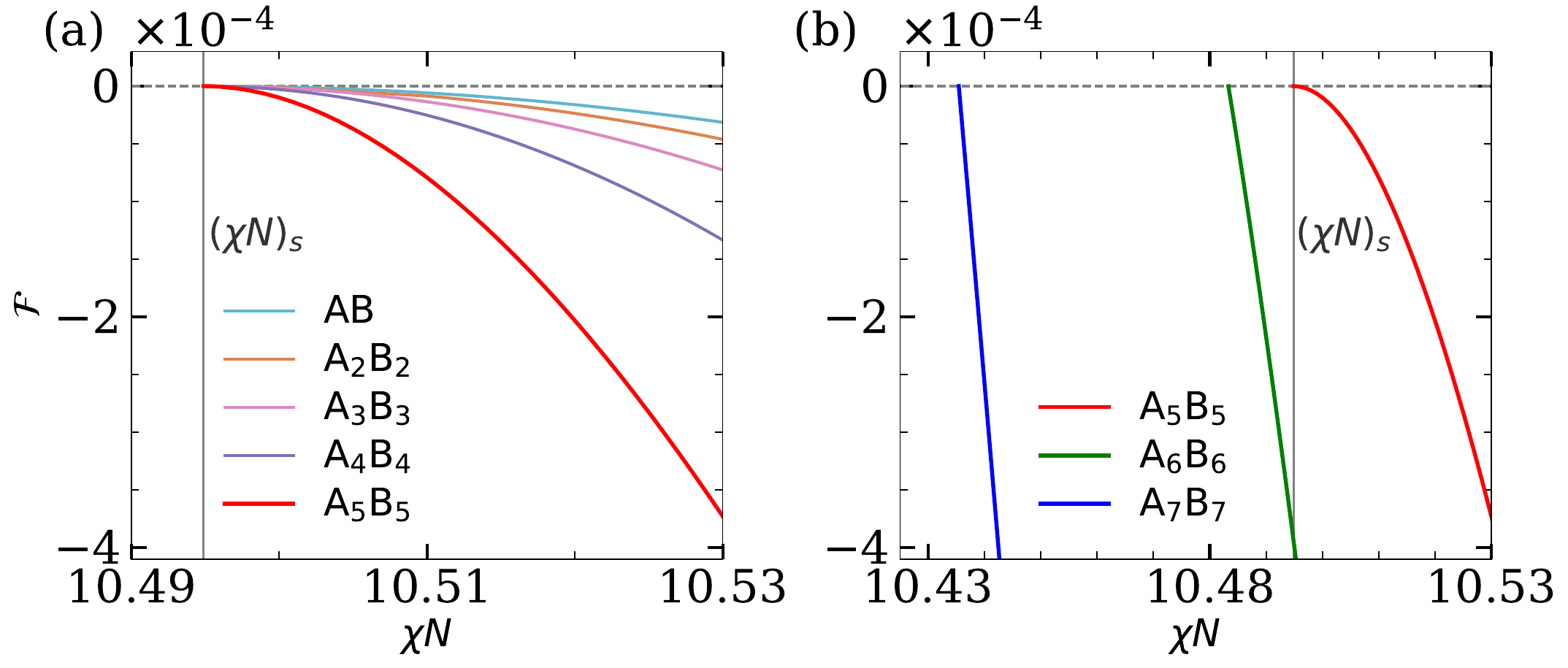}
\caption{Phase-transition behavior of symmetric \(\mathrm{A}_n\mathrm{B}_n\) star polymer systems:
(a) Second-order transitions for \(n \le 5\).
(b) First-order transitions for \(n \ge 6\).
The vertical gray line marks \((\chi N)_s \approx 10.495\).
}
\label{fig:figure1}
\end{figure}

To investigate the mean-field phase behavior, we performed standard SCFT calculations for continuous Gaussian chains (see Supplemental Material \cite{Supplemental}) by applying the saddle-point approximation to the free energy functional. All SCFT calculations were carried out using a customized version of the open-source Polymer FTS Python library \cite{Yong2022,Yongdd_langevin,Yong2025}.
The resulting free energies of the lamellar phase for $\mathrm{A}_n\mathrm{B}_n$ star polymers with $n = 1\text{--}7$ are presented in Fig.~\ref{fig:figure1}. Strikingly, the nature of the ODT depends sensitively on the arm number $n$. As shown in Fig.~\ref{fig:figure1}(a), for $n \le 5$, the free-energy curves vanish continuously at the spinodal point, $(\chi N)_{\mathrm{s}}$, while becoming tangent to the zero-free-energy baseline. This behavior indicates a second-order phase transition. In contrast, for $n \ge 6$ (Fig.~\ref{fig:figure1}(b)), the free-energy curves no longer merge smoothly with the zero-free-energy baseline at the spinodal point, $(\chi N)_{\mathrm{s}}$. Instead, they intersect the zero line with a finite slope, indicating that the lamellar and disordered phases have equal free energy at $(\chi N)_{\mathrm{ODT}} < (\chi N)_{\mathrm{s}}$. The transition therefore becomes first order. Consistently, the density profiles at the ODT show a continuous vanishing of the lamellar modulation for $n \le 5$, but a finite modulation amplitude for $n \ge 6$, corresponding to a discontinuous jump of the order parameter. Representative profiles are provided in the Supplemental Material \cite{Supplemental}. 

To analyze this behavior more rigorously, we develop an RPA scheme that is particularly convenient for evaluating higher-order terms in the symmetric case, \(f=1/2\). Since the instability of interest is toward a lamellar state, it is sufficient to consider one-dimensional periodic modulations about the homogeneous disordered state, chosen here as \(W_+=W_-=0\). We take the lamellar modulation to be along the \(x\) direction, so that the fields depend only on \(x\), and denote the lamellar period by \(2\pi/k\). Under \(\mathrm{A}\leftrightarrow\mathrm{B}\), \(W_+\) is invariant whereas \(W_-\) changes sign. In the symmetric lamellar state, this exchange is equivalent to a half-period translation, \(\pi/k\). Hence, \(W_+(x+\pi/k)=W_+(x)\) and \(W_-(x+\pi/k)=-W_-(x)\). Choosing the origin at a symmetry point of the lamellar profile, we may therefore expand the fields as
\begin{equation}
\begin{aligned}
W_-(x) &= C_1 \cos(kx) + C_3 \cos(3kx) + \cdots,\\
W_+(x) &= D_2 \cos(2kx) + D_4 \cos(4kx) + \cdots.
\end{aligned}
\end{equation}

At the saddle point, $W_-=\chi N(\phi_{\mathrm{B}}-\phi_{\mathrm{A}})/2$, where $\phi_i(\mathbf{r})$ is the $i$-type segment density at position $\mathbf{r}$. 
Near the quadratic instability of the disordered phase, the lamellar modulation is dominated by the fundamental mode, so we retain only $C_1$ in $W_-$ as the primary order-parameter amplitude. The even modes in $W_+$ require separate consideration. Although additional even harmonics $D_{2i}$ can arise in principle, we have verified that the second harmonic $D_2$ alone is sufficient for a consistent sixth-order free-energy expansion. This order is required for the complete analysis of the disorder-to-lamella transition. We henceforth write $C_1\equiv C$ and $D_2\equiv D$. The role of the higher harmonics at each expansion order is discussed in the Supplemental Material \cite{Supplemental}. 
With this truncation, 
\begin{equation}
\begin{aligned}
W_\mathrm{A}(x) &= C \cos(kx) + D \cos(2kx),\\
W_\mathrm{B}(x) &= -C \cos(kx) + D \cos(2kx).
\end{aligned}
\end{equation}

To evaluate the total partition function analytically near the ODT, we expand the propagators as perturbation series $q(x, s) = \sum_{i=0}^{\infty} q^{(i)}(x, s)$ and $p(x, s) = \sum_{i=0}^{\infty} p^{(i)}(x, s)$. The zeroth-order solutions are trivially given by $q^{(0)}(x, s) = p^{(0)}(x, s)=1$. Substituting these series into the original MDEs yields a recursive hierarchy of linear differential equations, in which the $(i+1)$th order term is driven by the $i$th order propagator coupled to the potential field
\begin{equation}
\begin{aligned}
\frac{\partial q^{(i+1)}(x,s)}{\partial s} &= \frac{a^2 N}{6} \frac{\partial^2 q^{(i+1)}(x,s)}{\partial x^2} - W_\mathrm{A}(x) q^{(i)}(x,s), 
\end{aligned} 
\end{equation}
with initial conditions $q^{(i)}(x, 0) = \delta_{i0}$. The same procedure applies to $p^{(i)}(x, s)$.

The total partition function is then obtained by spatially averaging the product of the expanded propagators over one lamellar period:
\begin{equation} \label{eq:totalq}
\begin{aligned}
Q &= \frac{k}{2\pi} \int_0 ^{\frac{2\pi}{k}}
\left(\sum_{i=0} ^\infty q^{(i)}\left(x,\frac{1}{2}\right)\right)^n
\left(\sum_{j=0} ^{\infty} p^{(j)}\left(x,\frac{1}{2}\right)\right)^n dx .
\end{aligned}
\end{equation}
We then expand $-\ln Q$ consistently through sixth order. Combining this expansion with the field terms in $\mathcal{F}$, the RPA free energy can be written as a symmetry-allowed expansion in the two amplitudes $C$ and $D$. Retaining the terms required for a sixth-order expansion in the primary amplitude $C$, we obtain
\begin{equation} \label{eq:fcdn}
\begin{aligned}
\mathcal{F}(C,D;n) &\approx 
 f_{2,0} C^2 
 + \left(f_{4,0} C^4 + f_{2,1} C^2D + f_{0,2} D^2\right) \\
&\quad
 + \left(f_{6,0} C^6 + f_{4,1} C^4D + f_{2,2} C^2D^2\right).
\end{aligned}
\end{equation}

The self-consistent solution must satisfy the incompressibility constraint, which is enforced by variation with respect to $W_+$. 
For the lamellar phase, the self-consistency equation reduces to
$\delta \mathcal{F}/\delta W_+ =
\left.\partial \mathcal{F}/\partial D\right|_{D=D^*}=0$. After substituting $D^*$ into Eq.~\eqref{eq:fcdn}, the free energy expression can be written solely in terms of $C$:
\begin{equation}
\begin{aligned}
\mathcal{F}(C;n)\approx f_{2}C^2 +f_{4}C^4+f_{6}C^6.
\end{aligned} 
\end{equation}
See the Jupyter notebook file in the Supplemental Material \cite{Supplemental} for analytic expressions of the coefficients $q^{(i)}$ and $p^{(i)}$ up to sixth order, together with the resulting quantities $f_{i,j}$ and $f_i$.
\\

We first revisit the well-established result for the spinodal point $(\chi N)_{\text{s}}$. 
The quadratic coefficient in free energy $\mathcal{F}$ is given by
\begin{equation}
\begin{aligned}
f_2 = \frac{1}{2\chi N} - \left(\frac{1}{2\mu} - \frac{3-4 e^{-\mu/2}+e^{-\mu}}{2\mu^2} \right),
\end{aligned} 
\end{equation}
where $\mu \equiv a^2N k^2/6$. The expression in parentheses attains its global maximum at $\mu^* \approx 3.7852$. For subsequent analysis, we evaluate all the coefficients after fixing $\mu = \mu^*$.
When the quartic coefficient is positive, the condition $f_2 = 0$ identifies the spinodal point, $(\chi N)_{\mathrm{s}} \approx 10.495$, which is the continuous disorder-to-lamella transition point of symmetric AB diblock copolymers.

For symmetric $\mathrm{A}_n\mathrm{B}_n$ star polymers, one immediately notices that $f_2$ is independent of the arm number $n$, consistent with the result of Olvera de la Cruz and Sanchez \cite{delacruz1986}. However, this conclusion is valid only when the quartic coefficient $f_4$ remains positive. As illustrated in Fig.~\ref{fig:figure1}(b), the transition changes its character as $n$ increases. To determine the order of the phase transition rigorously and to locate the tricritical point, we must therefore examine the quartic and sextic coefficients of the free energy, with particular attention to their dependence on $n$.

\begin{figure}
\centering
\includegraphics[width=\linewidth]{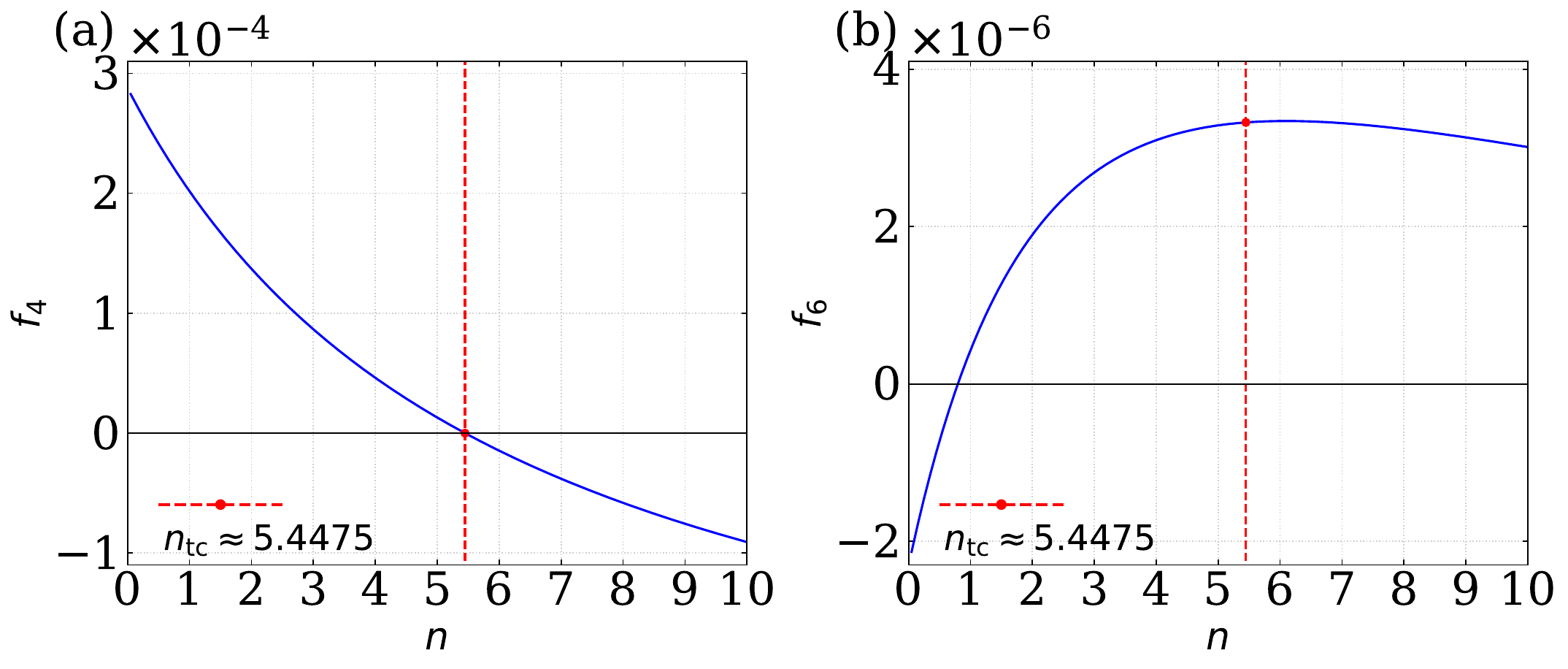}
\caption{Coefficients of \(\mathcal{F}\) as functions of \(n\):
(a) The quartic coefficient \(f_4\) vanishes and changes sign at \(n=n_{\mathrm{tc}} \approx 5.4475\).
(b) The sextic coefficient \(f_6\) remains positive in the vicinity of \(n=n_{\mathrm{tc}}\).
}
\label{fig:figure2}
\end{figure}

Our SCFT calculations (Fig.~\ref{fig:figure1}(b)) revealed a change in the transition character between $n = 5$ and $n = 6$ as predicted by Dobrynin and Erukhimovich~\cite{Dobrynin1991,Dobrynin1993}. In this sense, the negative quartic coefficient identified in their fourth-order theory already signaled a breakdown of that truncation as a complete description of the transition. To locate this crossover more precisely, we analytically extend our SCFT and RPA formalism by treating $n$ as a continuous positive real variable. Remarkably, we find that the quartic coefficient $f_4$ changes sign at a tricritical arm number $n_{\text{tc}} \approx 5.4475$, while the sextic coefficient $f_6$ remains strictly positive in its vicinity, as shown in Fig.~\ref{fig:figure2}. For $n < n_{\text{tc}}$, the quartic coefficient remains positive, leading to a continuous second-order transition at $(\chi N)_{\text{ODT}} = (\chi N)_{\text{s}}$. By contrast, for $n > n_{\text{tc}}$, the quartic coefficient becomes negative, dictating a first-order transition stabilized by the positive sextic coefficient. Although the sextic coefficient formally crosses zero again at $n\approx47$, this additional root lies outside the controlled regime of our small-parameter expansion and therefore does not have quantitative significance.

Furthermore, our sixth-order RPA calculation allows us to obtain an analytic expression for the phase boundary, \((\chi N)_{\mathrm{ODT}}\), on the first-order side of the tricritical point, where \(\delta n \equiv n-n_{\mathrm{tc}} > 0\) is small. The saddle-point condition, $\delta \mathcal{F}/\delta W_{-} =
\left.\partial \mathcal{F}(C;n)/\partial C\right|_{C=C^*}=0$, determines the equilibrium amplitude \(C^*\) and hence the ordered-phase free energy, \(\mathcal{F}(C^*;n)\). The first-order transition is determined by the condition
\begin{equation} \label{eq:Fcstarn}
\mathcal{F}(C^*;n)
\approx
-\frac{\Delta (\chi N)_{\mathrm{ODT}}}{2(\chi N)_{\mathrm{s}}^2}{C^*}^2
+ f_4 {C^*}^4
+ f_6 {C^*}^6
=0,
\end{equation}
where \(\Delta(\chi N)_{\mathrm{ODT}} \equiv (\chi N)_{\mathrm{ODT}}-(\chi N)_{\mathrm{s}}\), and \(f_i\) and \(C^*\) are implicit functions of \(n\).

\begin{figure}
\centering
\includegraphics[width=0.8\linewidth]{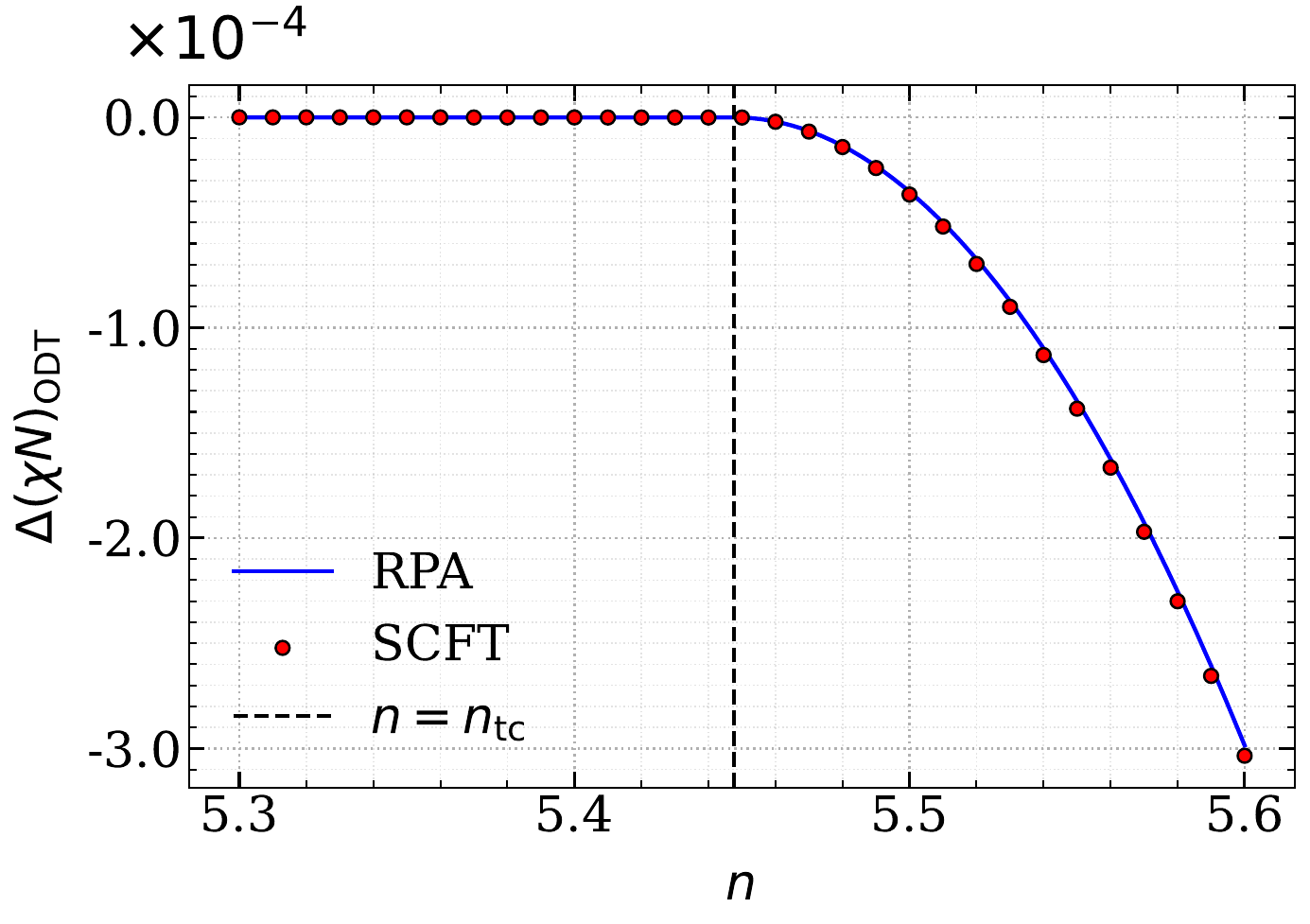}
\caption{
Comparison of the sixth-order RPA prediction for the phase boundary with the SCFT results.
}
\label{fig:figure3}
\end{figure}

Near the tricritical point, the coefficients \(f_4\) and \(f_6\) in Eq.~\eqref{eq:Fcstarn} inherit an implicit dependence on \(n\). In particular, because \(f_4(n_{\mathrm{tc}})=0\) at the tricritical point, the leading contribution on the first-order side is controlled by the linear variation of \(f_4\) with \(n\), which introduces \(f_4'(n_{\mathrm{tc}})\) into the phase-boundary shift. Carrying out the expansion for small \(\delta n \) then yields (see Supplemental Material \cite{Supplemental} for the full derivation)
\begin{equation}
\Delta (\chi N)_{\mathrm{ODT}} =
\left\{
\begin{array}{cl}
0, & n \le n_{\mathrm{tc}} \\[15pt]
\displaystyle
-\frac{(\chi N)_{\mathrm{s}}^2}{2}
\frac{\bigl(f_4^{\prime}(n_{\mathrm{tc}})\bigr)^2}{f_6(n_{\mathrm{tc}})}
\,\delta n^2,
& n > n_{\mathrm{tc}} .
\end{array}
\right.
\end{equation}
Thus, \(\Delta(\chi N)_{\mathrm{ODT}}\) is quadratic in \(\delta n\) near the tricritical point. Substituting the numerically evaluated coefficients at the tricritical point yields the explicit relation \(\Delta(\chi N)_{\mathrm{ODT}} \approx -0.012846\,\delta n^2\) for \(n > n_{\mathrm{tc}}\).
Figure \ref{fig:figure3} compares this analytic prediction with the SCFT results and shows remarkable quantitative agreement near the tricritical point, strongly supporting the validity of our theoretical framework.

Our sixth-order RPA formulation is sufficiently versatile to extend the analysis to mixtures, provided that the system is symmetric under the exchange of A and B. This extension provides an effective realization of the noninteger value of \(n_{\mathrm{tc}}\) through a binary mixture of symmetric star polymers, \(\mathrm{A}_{n_1}\mathrm{B}_{n_1}\) and \(\mathrm{A}_{n_2}\mathrm{B}_{n_2}\). Let \(\alpha\) denote the fraction of \(\mathrm{A}_{n_2}\mathrm{B}_{n_2}\) in the mixture. Since the uneliminated Landau coefficients \(f_{i,j}\) retained in the present expansion are affine functions of \(n\), the mixture free energy maps onto that of a monodisperse system with \(n_{\mathrm{eff}}=(1-\alpha)n_1+\alpha n_2\), or equivalently,
\begin{equation}
\begin{aligned}
\mathcal{F}_{\mathrm{mix}}(C,D)
&= (1-\alpha)\, \mathcal{F}(C,D;n_1)
 + \alpha\, \mathcal{F}(C,D;n_2) \\
&= \mathcal{F}\left(C,D;(1-\alpha)n_1+\alpha n_2\right) \\
&= \mathcal{F}(C,D;n_{\mathrm{eff}}).
\end{aligned}
\end{equation}
Choosing \(\alpha\) such that \(n_{\mathrm{eff}}=n_{\mathrm{tc}}\) therefore provides an effective realization of the tricritical condition in a binary mixture. An explicit comparison with SCFT calculations for an \(n_1=5\), \(n_2=6\) blend is given in the Supplemental Material~\cite{Supplemental}.

{\it Conclusion} --- We have developed an efficient higher-order free-energy expansion within the RPA framework to elucidate the tricritical behavior of $\mathrm{A}_n\mathrm{B}_n$ star polymer melts. 
Our sixth-order RPA analysis shows that $(\chi N)_{\mathrm{ODT}}$ departs from the spinodal point $(\chi N)_{\mathrm{s}}$ and that the transition becomes first order beyond a tricritical arm number $n_{\mathrm{tc}}$. These analytical predictions are in excellent agreement with SCFT calculations. Finally, we show that the noninteger value of $n_{\mathrm{tc}}$ admits an effective physical realization through an appropriate binary mixture of star polymers.

{\it Discussion} --- The origin of this tricritical behavior lies in junction-mediated inter-arm correlations in the symmetric star polymer ($f=1/2$). In the calculation of \((-\ln Q)/n\), selecting two factors among the \(q\) and \(p\) factors in Eq.~\eqref{eq:totalq} produces combinatorial factors such as \(n(n-1)/2\), which become contributions proportional to \(n-1\) after normalization by \(n\). These terms represent correlations between distinct arms sharing the common junction. In the quartic coefficient \(f_4\), the relevant junction-mediated contribution enters with a negative sign; as \(n\) increases, it drives \(f_4\) through zero at \(n_{\mathrm{tc}}\), while the sextic coefficient remains positive near the tricritical point. Thus, the emergence of a first-order ODT is not simply a consequence of star connectivity, but reflects the increasing weight of inter-arm correlations imposed by the common junction.

Although our analysis is formulated within a coarse-grained field-theoretic model based on Gaussian chains and local contact interactions, the result should be viewed as an asymptotically universal consequence of the \(\mathrm{A}_n\mathrm{B}_n\) star architecture in the long-chain limit, where microscopic chemical details enter only through renormalized parameters. At the same time, the Gaussian-arm approximation is expected to become less reliable at very large \(n\), where crowding near the junction can induce arm stretching and deviations from ideal chain statistics, so the large-arm-number limit should be interpreted with care. Moreover, both the sixth-order RPA theory and SCFT calculations are mean-field in nature, leaving fluctuation effects as an important open problem~\cite{Fredrickson1987,Mayes1991,Matsen2023}. Field-theoretic simulations provide a promising route for examining how fluctuations modify the phase boundary and the character of the ODT~\cite{Delaney2016,Matsen2021,Yong2022,Matsen2023}. More broadly, because the tricriticality identified here occurs at a finite ordering wavenumber and is controlled by star-polymer architecture, its critical behavior and universality class may differ from those of conventional local tricritical theories.

{\it Acknowledgments} --- This work was supported by National Research Foundation of Korea (NRF) grants funded by the Korea government (MSIT) (RS-2024-00348534 and 2022R1C1C2010613). We thank Prof. Mark Matsen of the University of Waterloo and Prof. June Huh of Korea University for helpful comments. This research used the high-performance computing resources of the UNIST Supercomputing Center. We acknowledge the use of ChatGPT as a coding assistant for the Python implementation and for light language editing of the manuscript. The Python scripts used to generate the results are provided in the Supplemental Material.

\bibliographystyle{apsrev4-2}
\bibliography{refs}      

\end{document}